\documentclass[onecolumn,preprintnumbers,amssymb,amsmath,superscriptaddress,letterpaper,nofootinbib]{revtex4}[12pt]

\usepackage{graphicx}
\usepackage{dcolumn}
\usepackage{bm}
\usepackage{natbib}
\usepackage{pstricks}
\usepackage{epsfig}
\usepackage{epstopdf}
\usepackage{slashed}

\newcommand{\be}{\begin{equation}}
	\newcommand{\ee}{\end{equation}}
\newcommand{\bea}{\begin{eqnarray}}
	\newcommand{\eea}{\end{eqnarray}}

\DeclareUnicodeCharacter{2009}{\,}

\begin{document}

	\title{ Spin-spin effects from non-relativistic limit of Dirac-Pauli-Maxwell Lagrangian}

	\author{Radmehr Fathi}
	\email{radmehr.fathi82@sharif.edu}
	\affiliation{Department of Physics, Sharif University of Technology, Tehran 11155-9161, Iran}
	
	\author{Nima Khosravi}
	\email{nima@sharif.edu}
	\affiliation{Department of Physics, Sharif University of Technology, Tehran 11155-9161, Iran}
	
	\date{\today}

	\begin{abstract}
		It is natural to expect that the electromagnetic spin-spin interaction has roots in a more fundamental theory: quantum electrodynamics in this case. To show this, we take the non-relativistic limit of the Dirac-Maxwell Lagrangian, using the Foldy-Wouthuysen transformation and derive the Breit interaction by an alternative approach. By adding the Pauli term (i.e., a higher-order interaction between spin and the electromagnetic field), we calculate the correction terms to electromagnetic spin-spin interactions in the non-relativistic limit. Finally, we study some of the physical effects of these terms regarding spin-spin interactions and spin-spin entanglements. Accordingly, we provide theoretical estimates that are not yet accessible with current experiments.
		
	\end{abstract}
	
	\maketitle

	\section{Introduction}
	The spin-spin interaction is well-known physics which is used extensively in condensed matter, statistical physics and quantum entanglement setups \cite{ZOUHAIR2014306,Ghosh:2021nqd,Hurst_2017,Amico:2007ag,Turek21042006,ABDALLA20123578,RevModPhys.97.025005,PhysRev.99.317}. This interaction should be predictable from fundamental physics. In this case, quantum electrodynamics (QED) should be responsible, and this effect has been examined by using the effective Hamiltonian from the scattering amplitude of two electrons in QED in \cite{PhysRev.34.553,PhysRev.39.616}. QED captures the physics of spin-$\frac{1}{2}$ particles and their interaction mediators (i.e., photons) by having the Dirac and the Maxwell Lagrangians in the same place. Although it seems natural to get the electromagnetic spin-spin interaction from the Dirac-Maxwell Lagrangian\footnote{It will be clear why we use the Dirac-Maxwell Lagrangian instead of QED very soon.} it has a nontrivial procedure. The Dirac-Maxwell Lagrangian is a relativistic theory, while the electromagnetic spin-spin interaction is a non-relativistic phenomenon. To connect these two, we need to take the non-relativistic limit of the Dirac-Maxwell Lagrangian. To do this, there is the well-known Foldy-Wouthuysen transformation \cite{PhysRev.78.29,PhysRev.111.1011,Silenko_2003,Haouam:2024hxu,Goncalves:2023zpc}, which is the standard way to take the non-relativistic limit of a spin system.
	
	It is important to clarify the details of the spin-photon-spin interaction. Of course, the Dirac-Maxwell Lagrangian produces such an interaction, but it is not the only one. If we write an effective field theory for this purpose, the Dirac-Maxwell Lagrangian (for this discussion ``$\bar\psi\,A_\mu\,\psi$") is the first term in the series. But there are higher order terms available, especially the Pauli term which is a non-minimal interaction, ``$\bar\psi\,F_{\mu\nu}\sigma^{\mu \nu}\,\psi$" \cite{Pauli1941original,PhysRev.73.416,Sastry:1999is,Muong-2:2025xyk,Aoyama_2020,Schwinger:1948iu}, with an additional derivative in comparison to ``$\bar\psi\,A_\mu\,\psi$". The additional derivative makes this term more important at higher energies while it can be sub-dominant at lower energies. The Pauli term was first introduced by Pauli to address the anomalous magnetic moment of the electron and other spin-half particles. This term can be generated exactly by calculating loop corrections in QED, which, in principle, is a UV-complete formulation of these kinds of interactions. 
	
	In the next section, we derive the Breit equation but with an alternative formalism, which is to treat one particle as the source of the magnetic field and the other one as the probe, introduce the magnetic field into the Pauli equation, and then restore the quantum nature of the spin of the source particle. Then in section \ref{dpm}, we introduce the Pauli term and study its non-relativistic limit in section \ref{fw} by employing the Foldy-Wouthuysen transformation. In section \ref{h}, we write the result induced by the Pauli term using our alternative approach in the Heisenberg-like form and study its physical effects in section \ref{ent}. We close the paper by the conclusion section. In addition, we have two appendices for details of magnetic field calculations and the Foldy-Wouthuysen transformation.

	\section{The Breit interaction from non-relativistic limit of Dirac-Maxwell Lagrangian}
	In this section, we will derive the Breit interaction from the non-relativistic limit of the Dirac-Maxwell Lagrangian. 
	The Dirac-Maxwell Lagrangian is:
	\begin{eqnarray}\label{eq:dirac-max}
		\mathcal{L}=-\frac{1}{4}F_{\mu \nu}F^{\mu \nu}+\bar{\psi}(i \slashed{D}-m) \psi
	\end{eqnarray}
	where $\slashed{X}\equiv\gamma^\mu X_\mu$ for Dirac matrices $\gamma^\mu$, $F_{\mu\nu}=\partial_\mu A_\nu-\partial_\nu A_\mu$, $A_\mu$ represents the electromagnetic four-vector and $\psi$ is the Dirac spinor. If we take the non-relativistic limit of this equation and keep the terms only to the order of $\frac{1}{m}$, we would get the Pauli equation
	\begin{eqnarray}
		i \, \frac{\partial \chi}{\partial t}
		= \left( m - e \varphi + \frac{\pi^2}{2 m}+ \frac{e}{2 m} (\vec{\sigma} \cdot \vec{B}) \right) \chi.
	\end{eqnarray}
	In order to derive the effective interaction between two identical electrons, we assume that the single spinor field $\psi$ consists of two separate fields $\psi_1$ , $\psi_2$. With this approach, we can treat one particle as the source of the electromagnetic field acting on the other. Thus the Lagrangian for our system can be written as:
	\begin{eqnarray}\label{eq:dirac-max-2}
		\mathcal{L}=-\frac{1}{4}F_{\mu \nu}F^{\mu \nu}+\bar{\psi}_1(i \slashed{D}-m) \psi_1+\bar{\psi}_2(i \slashed{D}-m) \psi_2.
	\end{eqnarray}
	It is important to emphasize that, at first glance, the above calculations seem to contradict the Pauli exclusion principle if $\psi_1$ and $\psi_2$ refer to the same type of particle, e.g., both refer to electrons.
	By introducing this effective two-field approach, the calculations become much simpler. However, it would cause some problems. By using this method, the Pauli exclusion principle will be lost. But we can fix that by forcing the two-fermion wave functions to be antisymmetric. Another problem is that if we do not force the wave functions to be antisymmetric, we lose the exchange energy. For the scattering amplitude of the two-field Lagrangian, we have:
	\begin{eqnarray}
		S=(-ie)^2
		\int d^4x_1\, d^4x_2\,
		T\!\left\{
		\bigl(\bar{\psi}_1(x_1)\gamma^\mu\psi_1(x_1)\bigr)
		\bigl(\bar{\psi}_2(x_2)\gamma^\nu\psi_2(x_2)\bigr)
		A_\mu(x_1)A_\nu(x_2)
		\right\}
	\end{eqnarray}
	we note that in general since we cannot contract $\psi_1$ and $\psi_2$, we can only take the direct type coupling from this scattering amplitude. However, if we calculate the quantum potential between the two fermions and then force the wave function to be antisymmetric, we would get both the direct and the exchange type of coupling. We note that this approach would only work in the non-relativistic limit, and in relativistic scenarios we would need to fully calculate everything based on quantum field theory. We note that this approach would induce another flavor for the electron. However, we do not need to worry about this second flavor since this is not a fundamental Lagrangian and it is just a mathematical method for simplifying the calculations.
	\\
	So now the corresponding equations of motion take the following form
	\begin{eqnarray}\label{eq:eq-motion}
		\partial_\mu\,F^{\mu \nu}=-(e \bar{\psi_1} \gamma^\nu \psi_1+e \bar{\psi_2} \gamma^\nu \psi_2).
	\end{eqnarray}
	So now, if we want to calculate the magnetic field sourced by the second particle (details of the calculation can be found in Appendix A), and if we only keep the spin terms, we would get:
	\begin{equation}
		\vec{B} (r_1)=\int d^3r_2\Big[\frac{-e}{8 m \pi} [\frac{3(\vec{r}_1-\vec{r}_2)((\chi_2^{\dagger} (r_2)\vec{\sigma}_2\chi_2 (r_2))\cdot(\vec{r}_1-\vec{r}_2))}{|\vec{r}_1-\vec{r}_2|^5}-\frac{(\chi_2^{\dagger} (r_2)\vec{\sigma}_2\chi_2 (r_2))}{|\vec{r}_1-\vec{r}_2|^3}]-\frac{e}{3 m} \delta^3 (\vec{r}_1-\vec{r}_2)(\chi_2^{\dagger} (r_2)\vec{\sigma}_2\chi_2 (r_2))\Big]
	\end{equation}
	as stated in Appendix A, $\chi_2$ is the wave function of the second particle in coordinate space as a function of $r_2$.\\
	So now if we put our magnetic field into the Pauli equation and only keep the spin-spin coupling part and then trace out the spatial part of the interaction, we would get:
	\begin{equation}
		\int d^3r_1d^3r_2\Big[\frac{-e^2}{16 m^2 \pi} [\frac{3((\vec{\alpha}_1)\cdot(\vec{r}_1-\vec{r}_2))((\vec{\alpha}_2)\cdot(\vec{r}_1-\vec{r}_2))}{|\vec{r}_1-\vec{r}_2|^5}-\frac{\vec{\alpha}_1\cdot \vec{\alpha}_2}{|\vec{r}_1-\vec{r}_2|^3}]-\frac{e^2}{6 m^2} \delta^3 (\vec{r}_1-\vec{r}_2)(\vec{\alpha}_1 \cdot \vec{\alpha}_2)\Big]
	\end{equation}
	where $\vec{\alpha}_i$ is defined as $\vec{\alpha}_i:=\chi_i^{\dagger } (r_i)\vec{\sigma}_i \chi_i (r_i)$. We note that this interaction is the same as the Breit interaction.
	\section{The electromagnetic field of the non-minimal coupling: Dirac-Pauli-Maxwell Lagrangian}\label{dpm}
	In this section, we examine the effects of corrections to the Dirac-Maxwell theory. These corrections, first introduced by Pauli \cite{Pauli1941original}, are crucial to predict the anomalous magnetic moment of electrons correctly. However, these corrections arise from loop corrections in QED. The Pauli term Lagrangian is:
	\begin{eqnarray}	
		\mathcal{L}_{\text{Pauli}}=-\frac{e \kappa}{2 m} \bar{\psi} \sigma^{\mu \nu} F_{\mu \nu} \psi
	\end{eqnarray}
	this yields the Dirac-Pauli-Maxwell Lagrangian where $\sigma^{\mu \nu}=\frac{i}{2}[\gamma^\mu,\gamma^\nu]$.  This new term can be seen as a higher-order interaction between electromagnetism and the spinors. However, this term arises when one calculates the loop corrections in quantum electrodynamics (QED). So $\kappa$ is a free phenomenological constant for the former and is a calculated one for the latter. In order to derive the effective interaction between two identical electrons, we again assume that the single spinor field $\psi$ consists of two separate fields $\psi_1$,$\psi_2$. With this approach, we can treat one particle as the source of the electromagnetic field acting on the other. Thus, the Lagrangian for our system can be written as:
	\begin{eqnarray}\label{eq:dirac-max-2}
		\mathcal{L}=-\frac{1}{4}F_{\mu \nu}F^{\mu \nu}+\bar{\psi}_1(i \slashed{D}-m) \psi_1-\frac{e \kappa}{2 m} \bar{\psi}_1 \sigma^{\mu \nu} F_{\mu \nu} \psi_1+\bar{\psi}_2(i \slashed{D}-m) \psi_2-\frac{e \kappa}{2 m} \bar{\psi}_2 \sigma^{\mu \nu} F_{\mu \nu} \psi_2.
	\end{eqnarray}
	The corresponding equations of motion take the following form 
	\begin{eqnarray}\label{eq:eq-motion-2}
		\partial_\mu\,F^{\mu \nu}+\frac{e \kappa}{m} \partial_\mu(\bar{\psi_1} \sigma^{\mu \nu} \psi_1+\bar{\psi_2} \sigma^{\mu \nu} \psi_2)=-(e \bar{\psi_1} \gamma^\nu \psi_1+e \bar{\psi_2} \gamma^\nu \psi_2)
	\end{eqnarray}
	for $A_\mu$ and
	\begin{eqnarray}\label{eq:dirac-eq-2}
		\left(\left(i\slashed{\partial}+e \slashed{A}-\frac{e \kappa}{2 m} \sigma^{\mu \nu} F_{\mu \nu}\right)-m\right) \psi_1=0
	\end{eqnarray}
	for $\psi_1$ and similarly for $\psi_2$. 
	
	Now our strategy is to treat one particle as the source of the electromagnetic field seen by the other particle. For this purpose, we focus on that part of the electromagnetic field which is sourced by one of the particles, for example the $\psi_1$, i.e. $\partial_\mu(F^{\mu \nu})=-e \bar{\psi_1} \gamma^\nu \psi_1+\frac{e \kappa}{m} \partial_\mu(\bar{\psi_1} \sigma^{\mu \nu} \psi_1)$. In the following, without confusion, we drop the index from $\psi_1$ and bring it back in the end. We also assume that the effect of the back reaction of the particles is negligible. We write the spinor as $\psi=\binom{\chi}{\xi}$ and consequently we have $\bar{\psi}=\psi^\dagger \gamma^0=(\chi^\dagger,  -\xi^\dagger )$. 
	Now we use the superposition principle and assume that the electromagnetic field has two different parts. One induced by the minimal coupling term and the other one induced by the Pauli term (non-minimal coupling),
	$F^{\mu \nu}=F^{\mu \nu}_{\text{minimal}}+F^{\mu \nu}_{\text{non-minimal}}$, and we have
	\begin{eqnarray}	
		D_\mu(F^{\mu \nu}_{\text{minimal}})=-(e \bar{\psi} \gamma^\nu \psi)
	\end{eqnarray}
	and
	\begin{eqnarray}	
		D_\mu(F^{\mu \nu}_{\text{non-minimal}})=\frac{-e \kappa}{m} D_\mu(\bar{\psi} \sigma^{\mu \nu} \psi).
	\end{eqnarray}
	The minimal part of the electromagnetic field was calculated in the previous section and for the non-minimal part of the electromagnetic field we have that according to $\sigma^{\mu \nu}=\frac{i}{2}[\gamma^\mu,\gamma^\nu]$:
	\begin{eqnarray}	
		\sigma^{i j}=\left(\begin{array}{cc}
			\epsilon_{i j k}\sigma^k & 0 \\
			0 & \epsilon_{i j k}\sigma^k
		\end{array}\right) .
	\end{eqnarray}
	So for the non-minimal part of the magnetic field, we have:
	\begin{eqnarray}\label{eq:curl}
		\partial_i F^{i j}_{\text{non-minimal}}=\epsilon_{i j k} \partial_i B_{k}^{\text{non-minimal}} &\simeq& \frac{-e \kappa}{m} \epsilon_{i j k} \partial_i(\chi^\dagger \sigma_k \chi) \\ \,\,\,\Rightarrow \,\,\, \vec{\nabla} \times \vec{B}^{\text{non-minimal}}&=&\vec{\nabla} \times (\frac{-e \kappa}{m}(\chi^\dagger \vec{\sigma} \chi)).
	\end{eqnarray}
	We also have that:
	\begin{eqnarray}\label{eq:div}
		\vec{\nabla} \cdot \vec{B}^{\text{non-minimal}}=0.
	\end{eqnarray}
	Now if we use Fourier transformation in order to solve equations (\ref{eq:curl}) and (\ref{eq:div}) together, we would have that:
	\begin{eqnarray}
		-k^2 \vec{\tilde{B}}^{\text{non-minimal}}=\vec{k} (\vec{k} \cdot (\vec{\tilde{b}}))-k^2 (\vec{\tilde{b}})
	\end{eqnarray}
	where $\vec{b}:=(\frac{-e \kappa}{m}(\chi^\dagger \vec{\sigma} \chi))$. Now if we go back to coordinate space, we would have that:
	\begin{eqnarray}
		\vec{B}^{\text{non-minimal}} (r_1)&=&(\frac{-e \kappa}{m})\int \Big[\frac{1}{4 \pi} [\frac{3(\vec{r}_1-\vec{r}_2)((\chi_2^{\dagger} (r_2)\vec{\sigma}_2\chi_2 (r_2))\cdot(\vec{r}_1-\vec{r}_2))}{|\vec{r}_1-\vec{r}_2|^5}-\frac{(\chi_2^{\dagger} (r_2)\vec{\sigma}_2\chi_2 (r_2))}{|\vec{r}_1-\vec{r}_2|^3}] \nonumber \\
		&& +\frac{2}{3} \delta^3 (\vec{r}_1-\vec{r}_2)(\chi_2^{\dagger} (r_2)\vec{\sigma}_2\chi_2 (r_2))\Big] d^3r_2
	\end{eqnarray}
	which has the same form as the minimal part of the magnetic field but it has a $2 \kappa$ coefficient.
	\section{The non-relativistic limit of the Dirac equation with the Pauli term: the Foldy-Wouthuysen transformation}\label{fw}
	Here, we derive a Schrödinger-like equation for our spin-half particles by taking the non-relativistic limit of the Dirac equation with the Pauli term. For this, we use the Foldy-Wouthuysen transformation (\cite{PhysRev.78.29,PhysRev.111.1011,Silenko_2003,Murgu_a_2010}), starting with the Dirac equation (\ref{eq:dirac-eq-2}) in the presence of the Pauli term, we have:
	\begin{eqnarray}
		\bigg( i\slashed{\partial} + e \slashed{A} 
		- \frac{e\kappa}{m} 
		( \,\vec{\sigma} \cdot \vec{B} 
		- i\,\vec{\alpha} \cdot \vec{E} )
		- m \bigg) \psi = 0,
	\end{eqnarray} 
	where the electric and magnetic fields are reintroduced. In this equation by $\vec{\sigma}$ we mean $\vec{\sigma} \otimes I$ and by $\vec{\alpha}$ we mean $\gamma^0 \vec{\gamma}$.
	The above equation can be rewritten as:
	\begin{eqnarray}
		\bigg( i\partial_0 + e\varphi - e\,\vec{\alpha} \cdot \vec{A} 
		+ i\,\vec{\alpha} \cdot \vec{\nabla} 
		- \frac{e\kappa}{m} 
		( \gamma^0\,\vec{\sigma} \cdot \vec{B} 
		- i\,\vec{\gamma} \cdot \vec{E} )
		- \gamma^0 m \bigg) \psi = 0
	\end{eqnarray} 
	here $\varphi$ is the scalar electric potential. The equation above can be seen as a Schrödinger equation, $i\,\partial_0\,\psi=H\,\psi$, giving the corresponding Hamiltonian 
	\begin{eqnarray}\label{Hamil}
		H 
		=  \gamma^0 m + \vec{\alpha} \cdot \vec{\pi} - e\varphi	+ \frac{e\kappa}{m}
		( \gamma^0\,\vec{\sigma} \cdot \vec{B} 
		- i\,\vec{\gamma} \cdot \vec{E} )
	\end{eqnarray}  
	where we have introduced the momentum operator, $-i \vec{\nabla}+e \vec{A}=\vec{\pi}$. 	
	The Foldy-Wouthuysen (FW) procedure requires separating the Hamiltonian into diagonal and off-diagonal blocks \cite{PhysRev.78.29,Barducci_2009}, i.e., $\mathcal{E}$ and $\mathcal{G}$ respectively. For the above Hamiltonian
	\begin{eqnarray}
		\mathcal{E} = -e\varphi + \frac{e\kappa}{m} \, \vec{\sigma} \cdot \vec{B}
	\end{eqnarray}
	and
	\begin{eqnarray}
		\mathcal{G} = \vec{\alpha} \cdot \vec{\pi} - \frac{i e \kappa}{m} (\vec{\gamma} \cdot \vec{E}).
	\end{eqnarray}
	To proceed in the FW procedure, the matrices should satisfy the following (anti-)commutation relations
	\begin{eqnarray}
		\mathcal{E} \gamma^0 = \gamma^0 \mathcal{E}, \quad \gamma^0 \mathcal{G} = -\mathcal{G} \gamma^0
	\end{eqnarray}
	which are satisfied in our case. There exists a matrix $S$ which transforms the Hamiltonian to a block-diagonal matrix
	\begin{eqnarray}
		S = - \frac{i}{2m} \gamma^0 \mathcal{G}=-\frac{i}{2 m}\bigg(\vec{\gamma} \cdot \vec{\pi}
		- \frac{i e \kappa}{m} \, (\vec{\alpha} \cdot \vec{E})\bigg).
	\end{eqnarray}
	The block-diagonal Hamiltonian, $H'$, takes the following form \cite{PhysRev.111.1011,Itzykson:1980rh,Wienczek_2022}
	\begin{eqnarray}\nonumber
		H' &=& H + i \, [S, H] 
		- \frac{1}{2} \, [S, [S, H]] 
		- \frac{i}{6} \, [S, [S, [S, H]]] 
		+ \frac{1}{24} \, [S, [S, [S, [S, H]]]] + \ldots \\
		&-& \dot{S} 
		- \frac{i}{2} \, [S, \dot{S}] 
		+ \frac{1}{6} \, [S, [S, \dot{S}]] + \ldots .
	\end{eqnarray}
	It is worth mentioning that for the case of a spin-half particle without the Pauli term ($\kappa=0$, i.e., $g=2$), the squared Dirac and Pauli operators show a supersymmetric structure. Which allows one to derive the exact, closed-form FW transformations under certain conditions \cite{Murgu_a_2010}. However, if we include the Pauli term in our Lagrangian ($\kappa\neq0$, i.e., $g\neq2$), this supersymmetry would break, and we can no longer derive an exact closed-form FW transformation. As a result, it is necessary to deploy a perturbative expansion in powers of $\frac{1}{m}$ \cite{Wienczek_2022,Barducci_2009} as we did here. It is worth noting that \cite{Barducci_2009} obtained closed forms of the Foldy-Wouthuysen transformation for some cases with an anomalous moment.
	\\
	For our Hamiltonian (\ref{Hamil}) the block diagonal Hamiltonian will be:
	\begin{eqnarray}\label{eq:hprime}
		H^{\prime}= \gamma^0 \left( m + \frac{1}{2 m} \, \mathcal{G}^2 \right)
		+ \mathcal{E} 
		- \frac{1}{8 m^2} \bigg[ \mathcal{G},  [\mathcal{G}, \mathcal{E} ] + i \, \dot{\mathcal{G}} \bigg].
	\end{eqnarray}
	For our purposes we are only keeping things to the terms up to ${\cal{O}}(\frac{1}{m^2})$ in the following.
	By using $\vec{F}_\pm=\vec{\pi} \mp \frac{i e \kappa}{m} \vec{E}$ we can rewrite the $\mathcal{G}$ matrix as 
	\begin{eqnarray}
		\mathcal{G}= 
		\begin{pmatrix}
			0 & \vec{\sigma} \cdot \vec{F}_+ \\
			\vec{\sigma} \cdot \vec{F}_- & 0
		\end{pmatrix}.
	\end{eqnarray}
	Since $\mathcal{G}^2$ is what appears in equation (\ref{eq:hprime}), it must be calculated explicitly. Since we need to find the matter part of the block-diagonal Hamiltonian, we need to find  $[ \mathcal{G}^2 ]^{1 1}$.
	Now, based on the details given in Appendix B, the ``$11$"-block of the $\mathcal{G}^2$ matrix and the rest of the terms that appear in equation (\ref{eq:hprime}) are: 
	\begin{eqnarray}
		[ \mathcal{G}^2 ]^{1 1} = \pi^2 + \frac{e \kappa}{m} \, (\vec{\nabla} \cdot \vec{E})
		+ e \, \vec{\sigma} \cdot \vec{B}
		+ \frac{e \kappa}{m} \, \vec{\sigma} \cdot 
		\left( 2 \, \vec{E} \times \vec{\pi} + i \, \vec{\nabla} \times \vec{E} \right)
	\end{eqnarray}
	and
	\begin{eqnarray}
		\bigg[ \mathcal{G},  [\mathcal{G}, \mathcal{E} ] + i \, \dot{\mathcal{G}} \bigg]
		= i e \, \vec{\sigma} \cdot ((\vec{\nabla} \times \vec{E}) 
		+ 2 e \, (\vec{E} \times \vec{\pi}))+e (\vec{\nabla} \cdot \vec{E}).
	\end{eqnarray}
	So based on these calculations the non-relativistic limit of the Dirac equation in the presence of the Pauli term to the order of $\frac{1}{m^2}$ would be:
	
	\begin{eqnarray}\label{eq:FWPauli}\nonumber
		i \, \frac{\partial \chi}{\partial t}
		&=& \left( m - e \varphi + \frac{1}{2 m} \left(\pi^2 
		\right)+\frac{e}{8 m^2} \left( \vec{\nabla} \cdot \vec{E} 
		+ i \vec{\sigma} \cdot (\vec{\nabla} \times \vec{E}) 
		+ 2 \, \vec{\sigma} \cdot (\vec{E} \times \vec{\pi}) \right) \right) \chi\\
		&+& \frac{e \kappa}{2 m^2} \left((\vec{\nabla} \cdot \vec{E})
		+ \vec{\sigma} \cdot (i (\vec{\nabla} \times \vec{E}) 
		+ 2 (\vec{E} \times \vec{\pi}))\right) \chi+ \frac{e}{2 m}(1+2 \kappa) (\vec{\sigma} \cdot \vec{B})\chi,
	\end{eqnarray}
	where $\chi$ is the matter part of the wave function $\psi=\binom{\chi}{\xi}$.
	
	\section{A Heisenberg-like interaction induced by the Pauli term}\label{h}
	So now we use the dominant term in our non-relativistic limit of the Dirac equation with the Pauli term ($\frac{e}{2 m} (1+2 \kappa) \, (\vec{\sigma} \cdot \vec{B})$). Then we use the contact term in the non-minimal part of the magnetic field $\vec{B}$ that we calculated. We would get a contact term induced by the Pauli term:
	\begin{eqnarray}
		H_1 \chi_1=\bigg(\frac{-e^2 \kappa}{3 m^2}\bigg)\times\bigg(1+2 \kappa\bigg)\times\bigg(\vec{\sigma} \cdot (\chi_2^\dagger \vec{\sigma} \chi_2)\bigg) \chi_1=\epsilon \chi_1.
	\end{eqnarray}
	Now if we write $\chi_1$
	as a product of the spatial part and the spin part, $\chi= \psi_{(\vec{r})} \phi_{\text{spin}}$, and then integrate out the spatial part, by multiplying $\psi^\dagger_1$ on both sides and then integrating, we would have:
	\begin{eqnarray}	
		\int H_1 \psi_1^{\dagger} \psi_1 \phi_{1} d^3 r=\bigg[
		\bigg(\frac{-e^2 \kappa}{3 m^2}\bigg)\times\bigg(1+2 \kappa\bigg)\times\bigg(\vec{\sigma} \cdot (\phi_2^\dagger \vec{\sigma} \phi_2)\bigg)\times\bigg (\int \psi_1^{\dagger} \psi_1 \psi_2^{\dagger} \psi_2  d^3 r\bigg) \bigg]\phi_1.
	\end{eqnarray}
	Now for the spin-spin interacting Hamiltonian, $\hat{H}$, we have:
	\begin{eqnarray}
		\boxed{\hat{H} = \bigg(\frac{-e^2 \, w}{3\, m^2}\,\,\kappa\,\,(1+2 \kappa)\bigg)\,\,\vec{\sigma} \otimes \vec{\sigma}}
	\end{eqnarray}
	where we defined $w=\int \psi_1^{\dagger} \psi_1 \psi_2^{\dagger} \psi_2 d^3 r$. Then we promoted the expectation values to quantum operators in the non-minimal part of the magnetic field. The result is the electromagnetic spin-spin interaction as the Hamiltonian of our composite system, which has the same form as the Heisenberg Hamiltonian. This shows that if we assume the Pauli term ($\kappa\neq 0$) in our Lagrangian, we would get a contact term induced by the Pauli term in addition to the Breit interaction, and the contact term is only non-zero if the two particle wave functions overlap with each other ($w\neq 0$). We also note that because of the Pauli term and because of the $(1+2 \kappa)$ coefficient in $\frac{e}{2 m}(1+2 \kappa) (\vec{\sigma} \cdot \vec{B})$, the Breit interaction will be multiplied by a factor of $(1+2 \kappa)$.

	\section{Physical Effects}\label{ent}
	It is instructive to see if the above results can be observed by a laboratory experiment. For our own purposes\footnote{Our main goal for this work is its generalization for the case of the gravitational field, which is a work in progress. There we follow the approach in \cite{Marletto_2017,Bose_2017}.}, we want to examine this effect in the entanglement between two spins. If we examine a two-electron system, the effect induced by the Pauli term would be small compared to the Breit interaction. However, if we examine other spin-half particles like protons, the contribution becomes more significant. Finally, for neutrons, the Breit interaction becomes negligible, and the non-minimal part of the electromagnetic field dominates. We note that neutrons are neutral and, because of that, have no minimal coupling to the electromagnetic field. However, because of the quark structure of the neutron, there exists a non-minimal coupling in the form of the Pauli interaction. The effect induced by the Pauli term only exists if our particle wave functions overlap. In the Breit equation, if the wave functions overlap, then the interaction might diverge. However, we can still examine the contact term in both the Breit equation and the effect induced by the Pauli term. In cases where the angular part of the integral forces the $\frac{1}{r^3}$ term to become zero. Since neutrons have no electric charge, the Breit interaction is negligible for them. Now if we assume that the Pauli term for neutrons has the form \cite{Bruce:2020jlt,Steinmetz:2018ryf} $\mathcal{L}=-\frac{e \kappa_n}{2 M_n} \bar{\psi}F^{\mu \nu} \sigma_{\mu \nu}\psi$, then the main term of the interaction between two neutrons induced by the Pauli interaction would be 
	\begin{eqnarray}
		\hat{H}=\frac{-2 e^2 \kappa_n^2 w}{3 M_n^2} \vec{\sigma} \otimes \vec{\sigma}
	\end{eqnarray}
	where $\kappa_n=-0.96$ \cite{Bruce:2020jlt} and $M_n$ is the mass of the neutron. Similarly one could calculate the main term of the interaction between a neutron and a proton induced by the Pauli term, which would be:
	\begin{eqnarray}
		\hat{H}_{np}=\frac{- 2 e^2 \kappa_n \kappa_p w}{3 M_n M_p} \vec{\sigma} \otimes \vec{\sigma}
	\end{eqnarray}
	where $\kappa_p=0.896$ and $M_p$ is the mass of proton. We note that we are assuming that the Pauli term Lagrangian of the proton has the form $\mathcal{L}=-\frac{e\kappa_p}{2 M_p} \bar{\psi}F^{\mu \nu} \sigma_{\mu \nu}\psi$.
	If we assume that a system which consists of one neutron and one proton is in a separable state at $t=0$ and this system interacts only via our Hamiltonian, then observation of any entanglement means observation of the above Hamiltonian. Although the neutron-proton system is only a theoretical suggestion, the same formalism applies to more experimentally viable systems, such as the nuclear spin of two trapped neutral atoms.
	\\
	We assume that by trapping the neutron and the proton, we have set the spatial part of our particles' wave functions to Gaussian wave functions (which is a good assumption because the ground state of a simple harmonic oscillator is a Gaussian state). We also assume that we can set the spatial part of the wave functions without forcing any constraints on the spin part of the wave functions. For simplicity, we assume both particles have the same variance, $\sigma_r$. So we have:
	\begin{eqnarray}
		\psi_i^{\dagger} \psi_i = \frac{1}{(2 \pi \sigma_r^2)^{\frac{3}{2}}} e^{-\frac{(\vec{r}-\vec{\mu}_i)^2}{2 \sigma_r^2}}
	\end{eqnarray}
	where $\vec{\mu}_i$ represents the mean distance for each particle. Then we have
	\begin{eqnarray}
		w=\int \psi_1^{\dagger} \psi_1 \psi_2^{\dagger} \psi_2 d^3 r=\frac{1}{(4 \pi \sigma_r^2)^{\frac{3}{2}}} e^{-\frac{(\vec{\mu}_1-\vec{\mu}_2)^2}{4 \sigma_r^2}}.
	\end{eqnarray}	
	If our Hamiltonian has the form $\hat{H}= J (\vec{\sigma} \otimes \vec{\sigma})$ then we can calculate $J$: 
	\begin{eqnarray}
		J(\Delta r)=\bigg(\frac{-2 e^2\,\kappa_n \kappa_p }{3 M_n M_p}\bigg)\, \frac{1}{(4 \pi \sigma_r^2)^{\frac{3}{2}}} \,e^{-\frac{(\Delta r)^2}{4 \sigma_r^2}}
	\end{eqnarray}
	where $\Delta r=|\vec{\mu}_2-\vec{\mu}_1|$.
	Now if we assume that the two particles have the separable state $| 0 1 \rangle$ at $t=0$ (which does not break the Pauli exclusion principle, since we are not dealing with identical particles) and calculate the entanglement entropy which is equal to $S=-Tr(\rho_A \log(\rho_A))$
	after a time $t$ we would get:
	\begin{eqnarray}\nonumber
		S = - \cos^2(2 J t) \log{(\cos^2(2 J t))}
		- \sin^2(2 J t) \log{(\sin^2(2 J t))}.
	\end{eqnarray}
	It is easy to see that $S$ reaches its maximum value, $\log(2)$, for $J =\pi/(8 t)$. The decoherence time, which is usually very small for these kinds of systems, limits our freedom to play with the time parameter. This means $J$ is (effectively) the only freedom we have to play with. In the following, we try to see how one can get higher $J$ values to get observable values of $S$ in very small time intervals. We can set the wave function of our particles to whatever we can make in the lab. Still, we should be careful not to violate the non-relativistic limit (for example, if we assume that the $z$-dependence of our wave functions is a Dirac delta function, we would get infinity for the value of $w$). However, we cannot use this wave function since it would give infinite momentum uncertainty, which obviously violates the non-relativistic limit.)
	In the above examination, we assumed we could confine the neutrons in the lab and focused solely on the electromagnetic spin-spin interaction, without examining the strong-force interactions between the neutron and the proton. The strong interaction between a contacting neutron and proton dominates the interaction induced by the Pauli term. For this reason, our examination of entanglement serves only as a mathematical model. However, for more realistic models, we can examine the interaction between the nuclear spin of two neutral atoms. It is worth mentioning that in the case of the neutral atoms, all the magnetic moment comes from the Pauli term since the standard Dirac-Maxwell model does not produce any of it. For example, if we examine the interaction induced by the Pauli term between an atom of ${}^{171}$Yb and an atom of ${}^{3}$He, we would get $J$ as:

	\begin{eqnarray}
		J(\Delta r)=\bigg(\frac{-2 e^2 \,\kappa_{Yb} \kappa_{He} }{ 3 M_{Yb} M_{He}}\bigg)\, \frac{1}{(4 \pi \sigma_r^2)^{\frac{3}{2}}} \,e^{-\frac{(\Delta r)^2}{4 \sigma_r^2}}
	\end{eqnarray}
	where $\kappa_{Yb}=42.2$, $\kappa_{He}=-3.19$ \cite{STONE200575}, $M_{Yb}=159.3 (\text{GeV})$ and $M_{He}=2.8 (\text{GeV})$. Now by assuming $\sigma_r= 30 (\text{nm})$ and maximizing  the overlap between the states i.e. $\Delta r=0$, we obtain $|J(0)|\simeq 1.18 \times 10^{-19} (\text{eV})$. Note that this value is a theoretical estimation. We also note that observing the effect in the case where there is complete overlap of two neutral atom wave functions ($\Delta r=0$) is not straightforward.
	\\ 
	Now we would like to examine the alteration of the interaction between two electrons in a helium atom due to the Pauli term. We note that the Pauli term would alter the Breit interaction.
	For obtaining the altered Breit equation, we need to use the last term in equation (\ref{eq:FWPauli}) which is $\frac{e}{2 m}(1+2 \kappa) (\vec{\sigma} \cdot \vec{B})$ and then employ the magnetic field that contains both the minimal and the non-minimal parts:
	\begin{eqnarray}
		\vec{B} (r_1)&=&\int \Big[\frac{-e(1+2 \kappa)}{8 m \pi} [\frac{3(\vec{r}_1-\vec{r}_2)((\chi_2^{\dagger} (r_2)\vec{\sigma}_2\chi_2 (r_2))\cdot(\vec{r}_1-\vec{r}_2))}{|\vec{r}_1-\vec{r}_2|^5}-\frac{(\chi_2^{\dagger} (r_2)\vec{\sigma}_2\chi_2 (r_2))}{|\vec{r}_1-\vec{r}_2|^3}] \nonumber \\
		&-&\frac{e (1+2 \kappa)}{3 m} \delta^3 (\vec{r}_1-\vec{r}_2)(\chi_2^{\dagger} (r_2)\vec{\sigma}_2\chi_2 (r_2)) \Big] d^3r_2 .
	\end{eqnarray}
	Which results in the altered Breit interaction:
	\begin{equation}\label{eq:mod-Breit}
		(1+2 \kappa)^2\Big(\frac{-e^2}{16 m^2 \pi} \big[\frac{3(\vec{\alpha}_1\cdot\vec{r})(\vec{\alpha}_2\cdot\vec{r})}{r^5}-\frac{\vec{\alpha}_1\cdot \vec{\alpha}_2}{r^3}\big]-\frac{e^2}{6 m^2} \delta^3 (\vec{r})(\vec{\alpha}_1 \cdot \vec{\alpha}_2)\bigg).
	\end{equation}
	We note that, as expected, the altered Breit equation is the same as the standard Breit equation times $(1+2 \kappa)^2$. Now, if we use this modified interaction in order to calculate the
	interaction strength between two electrons in the $l=0$ state in the helium atom, our result would reproduce an
	established radiative correction \cite{PhysRev.99.317} which is $\frac{-2 e^2 (\kappa+\kappa^2) w}{3 m^2}$, to the second order of $\kappa$. This is in complete agreement with what is known. We note that this quantity can also be calculated for helium-like atoms. It is also worth noting that due to the $(1+2 \kappa)^2$ coefficient, equation (\ref{eq:mod-Breit}) predicts the amplitude of the interaction between two protons $(\kappa_p \simeq 0.896)$ to be approximately 7.8 times the amount that the standard Breit interaction predicts.
	\section{Concluding remarks}\label{con}
	In this work, we tried to see if the electromagnetic spin-spin, $J (\vec\sigma \cdot \vec\sigma)$ interaction in low-energy limits (e.g., in condensed matter physics) can be a probe of high-energy physics. The standard Dirac-Maxwell Lagrangian predicts such a term in its non-relativistic limit, which is a result given by Breit \cite{PhysRev.34.553,PhysRev.39.616}. However, in the language of effective field theory, we can consider additional terms to the Dirac-Maxwell Lagrangian\footnote{As we mentioned in the paper, these terms can be seen as loop corrections in the QED formalism or due to higher-order terms from new physics.}. For this purpose, we consider the Pauli term, which is a higher-order correction to the Dirac-Maxwell Lagrangian. To achieve this goal, we employed the Foldy-Wouthuysen approach to take the non-relativistic limit of our models. First, this method can derive the Breit equation by an alternative approach compared with the standard one \cite{PhysRev.34.553,PhysRev.39.616}. We then calculated the effects of the Pauli term, which affects only the interaction strength, $J$, not its form. We examined methods to observe the effects of additional terms and estimated theoretical values that may be detectable in future experiments. One way to examine these effects is by measuring the entanglement between two spins induced by this interaction. This work introduces a benchmark for comparison of non-relativistic limits of other theories, e.g., gravity \cite{FW-Gravity} and its extensions. This can help us to look for the quantum effects of gravity in the direction of \cite{Marletto_2017,Bose_2017}.
	We also note that this problem was examined using an effective approach in \cite{PhysRevB.81.184419}. In that work, they first took the non-relativistic limit of the Dirac equation and then constructed an action based on that limit. While their overall results are similar to ours, our methodology uses the Foldy-Wouthuysen transformation, which provides a more systematic and rigorous framework for handling higher-order corrections. Small differences arise because of the different structure in the Lagrangian and the difference in the methodology. This highlights the importance of using a rigorous approach for calculating higher-order terms.
	
	\section*{Appendix A: The magnetic field}\label{app:A}
	We define the Fourier transform of the field $A^\mu(x)$ and current $J^\mu(x)$ as:
	\begin{equation}
		A^\mu(x) = \int \frac{d^4k}{(2\pi)^4} \tilde{A}^\mu(k) e^{-ik\cdot x}, \quad
		J^\mu(x) = \int \frac{d^4k}{(2\pi)^4} \tilde{J}^\mu(k) e^{-ik\cdot x}.
	\end{equation}
	Thus, the vector potential in momentum space is:
	\begin{equation}
		\vec{\tilde{A}}(k) =\frac{1}{k^2} \vec{\tilde{J}}_\perp(k)
	\end{equation}
	where the transverse current is defined as
	$
	\vec{\tilde{J}}_\perp(k) = \vec{\tilde{J}}(k) - \vec{k} \frac{(\vec{k} \cdot \vec{\tilde{J}}(k))}{|\vec{k}|^2}
	$
	and the initial state field $\psi(x)$ that corresponds to momentum $p$ is defined as
	$
	\psi(x) = \int \frac{d^3p}{(2\pi)^3} u(\vec{p}) e^{-ip \cdot x}.
	$
	\\
	So the density current would become
	\begin{equation}
		J^\mu(x) = -e \int \frac{d^3p'}{(2\pi)^3} \int \frac{d^3p}{(2\pi)^3} \left[ \bar{u}(\vec{p}') \gamma^\mu u(\vec{p}) \right] e^{i(p' - p) \cdot x}
	\end{equation}
	and we define the Fourier transform of the current $\tilde{J}^\mu(k)$ as
	$
	\tilde{J}^\mu(k) = \int d^4x \, J^\mu(x) e^{ik \cdot x}.
	$
	Now if we substitute the expression for $J^\mu(x)$ into the integral we would get
	\begin{equation}
		\tilde{J}^\mu(k) = -\int d^4x \, e^{ik \cdot x} \left( e \int \frac{d^3p'}{(2\pi)^3} \int \frac{d^3p}{(2\pi)^3}  \bar{u}(\vec{p}') \gamma^\mu u(\vec{p}) e^{i(p' - p) \cdot x} \right)
	\end{equation}
	which equals
	\begin{equation}
		\tilde{J}^\mu(k) = -e \int \frac{d^3p}{(2\pi)^3}  \bar{u}(\vec{p}+\vec{k}) \gamma^\mu u(\vec{p})
	\end{equation}
	To simplify the current density $\tilde{J}^\mu(k)$, we apply the Foldy-Wouthuysen (FW) transformation approach by expanding the Dirac spinors in powers of $1/m$. This effectively decouples the large and small components.
	\begin{equation}
		u(\vec{p}) \approx \begin{pmatrix} \tilde{\chi} \\ \frac{\vec{\sigma} \cdot \vec{p}}{2m} \tilde{\chi} \end{pmatrix}
	\end{equation}
	then we would get
	\begin{align}
		u'^\dagger \vec{\alpha} u &\approx \begin{pmatrix} \tilde{\chi}'^\dagger & \tilde{\chi}'^\dagger \frac{\vec{\sigma}\cdot\vec{p}'}{2m} \end{pmatrix} 
		\begin{pmatrix} 0 & \vec{\sigma} \\ \vec{\sigma} & 0 \end{pmatrix} 
		\begin{pmatrix} \tilde{\chi} \\ \frac{\vec{\sigma}\cdot\vec{p}}{2m} \tilde{\chi} \end{pmatrix} \\
	\end{align}
	which equals
	\begin{equation}
		\frac{1}{2 m} \tilde{\chi}'^\dagger \left[ (\vec{\sigma}\cdot\vec{p}')\vec{\sigma} + \vec{\sigma}(\vec{\sigma}\cdot\vec{p}) \right] \tilde{\chi}
	\end{equation}
	which simplifies to
	\begin{equation}
		\bar{u}' \vec{\gamma} u \approx \frac{1}{2 m} \tilde{\chi}'^\dagger \left[ (2\vec{p} + \vec{k}) - i \vec{\sigma} \times \vec{k} \right] \tilde{\chi}.
	\end{equation}
	In momentum space the magnetic field is defined as
	\begin{equation}
		\vec{\tilde{B}}(k) = i\vec{k} \times \vec{\tilde{A}}(k)
	\end{equation}
	now if we substitute the expression for $\vec{\tilde{A}}(k)$
	\begin{equation}
		\vec{\tilde{B}}(k) = i\vec{k} \times \left[ \frac{1}{k^2} \left( \vec{\tilde{J}}(k) - \frac{\vec{k}(\vec{k} \cdot \vec{\tilde{J}})}{|\vec{k}|^2} \right) \right]
		=\frac{i}{k^2} (\vec{k} \times \vec{\tilde{J}}(k))
	\end{equation}
	now if we put one of our previous results back into this equation we would get
	\begin{equation}
		\vec{\tilde{B}}(k) = \frac{-ie}{2m k^2} \int \frac{d^3p}{(2\pi)^3}  \tilde{\chi}^{\dagger \prime} \left[ \vec{k} \times (2\vec{p}+\vec{k}) - \vec{k} \times (i\vec{\sigma} \times \vec{k}) \right] \tilde{\chi}
	\end{equation}
	which simplifies to
	\begin{equation}
		\vec{\tilde{B}}(k) = \frac{-ie}{2m k^2} \int \frac{d^3p}{(2\pi)^3}\tilde{\chi}^{\dagger \prime} \Big[ 2(\vec{k} \times \vec{p}) - i(k^2\vec{\sigma} - \vec{k}(\vec{k}\cdot\vec{\sigma})) \Big] \tilde{\chi}.
	\end{equation}
	So now we keep the spin part of the magnetic field and then write $\tilde{\chi}(p)=\int d^3x e^{i p \cdot x} \chi(x)$ and find the magnetic moment at the point $r_1$, which is the point where the first particle is. We will also give $\chi$ indices of 1 and 2 in order to separate the two particles.
	\begin{equation}
		\vec{B} (r_1)=\int d^3r_2\Big[\frac{-e}{8 m \pi} [\frac{3(\vec{r_1}-\vec{r_2})((\chi_2^{\dagger} (r_2)\vec{\sigma_2}\chi_2 (r_2))\cdot(\vec{r_1}-\vec{r_2}))}{|\vec{r_1}-\vec{r_2}|^5}-\frac{(\chi_2^{\dagger} (r_2)\vec{\sigma_2}\chi_2 (r_2))}{|\vec{r_1}-\vec{r_2}|^3}]-\frac{e}{3 m} \delta^3 (\vec{r_1}-\vec{r_2})(\chi_2^{\dagger} (r_2)\vec{\sigma_2}\chi_2 (r_2))\Big].
	\end{equation}

	\section*{Appendix B: The Foldy-Wouthuysen transformation}\label{app:B}
	We start by calculating what we need in order to calculate the diagonalized Hamiltonian. The $\mathcal{G}^2$ matrix has the following form
	\begin{eqnarray}
		\mathcal{G}^2 = 
		\begin{pmatrix}
			(\vec{\sigma} \cdot \vec{F}_+) \, (\vec{\sigma} \cdot \vec{F}_-) & 0 \\
			0 & (\vec{\sigma} \cdot \vec{F}_-) \, (\vec{\sigma} \cdot \vec{F}_+)
		\end{pmatrix}.
	\end{eqnarray}
	The "11"-component of the above matrix  can be derived by taking the following steps
	\begin{eqnarray}
		(\sigma_i \sigma_j) \, (F_{+ i} F_{-j}) 
		= (\vec{F}_{+} \cdot \vec{F}_{-})
		+ \, i \epsilon^{ijk} \sigma_k \, (F_{+i} F_{-j}) 
		= (\vec{F}_{+} \cdot \vec{F}_{-})
		+ \, i \, \vec{\sigma} \cdot (\vec{F_+} \times \vec{F_-}).
	\end{eqnarray}
	Now we can write the first term as
	\begin{eqnarray}
		\vec{F}_{+} \cdot \vec{F}_{-}=(\vec{\pi} - \frac{i e \kappa}{m} \vec{E})\cdot (\vec{\pi}+\frac{i e \kappa}{m} \vec{E})=
		\pi^2 + \frac{i e \kappa}{m} (\vec{\pi} \cdot \vec{E})
		+ \frac{e^2 \kappa^2}{m^2} E^2
		- \frac{i e \kappa}{m} (\vec{E} \cdot \vec{\pi}).
	\end{eqnarray}
	We also have
	\begin{eqnarray}
		(\vec{E} \cdot \vec{\pi} - \vec{\pi} \cdot \vec{E}) \, \psi 
		&=& (\vec{E} \cdot (-i\vec{\nabla} + e\vec{A}) - (-i\vec{\nabla} + e\vec{A}) \cdot \vec{E}) \, \psi \\ \nonumber
		&=& - e \, \vec{E} \cdot \vec{A} \, \psi
		-i \vec{E} \cdot (\vec{\nabla} \psi) + e \, \vec{A} \cdot \vec{E} \, \psi
		+ i \, \vec{\nabla} \cdot (\vec{E} \, \psi)
		= i \, (\vec{\nabla} \cdot \vec{E}) \, \psi.
	\end{eqnarray}
	So the first term in ``11"-component of $\mathcal{G}^2$ can be written as
	\begin{eqnarray}
		(\vec{\sigma} \cdot \vec{F}_{+}) \, (\vec{\sigma} \cdot \vec{F}_{-})
		= \pi^2 + \frac{e\kappa}{m} \, (\vec{\nabla} \cdot \vec{E})+\frac{e^2 \kappa^2}{m^2} \, E^2
		+ i \, \vec{\sigma} \cdot (\vec{F}_+ \times \vec{F}_{-}).
	\end{eqnarray}
	The last term can also be simplified by taking the following steps 
	\begin{eqnarray}
		\vec{F}_+ \times \vec{F}_{-}= (\vec{\pi} - \frac{i e \kappa}{m} \vec{E}) \times 
		(\vec{\pi} + \frac{i e \kappa}{m} \vec{E})&=& (\vec{\pi} \times \vec{\pi})
		- \frac{i e \kappa}{m} (\vec{E} \times \vec{\pi}-\vec{\pi} \times \vec{E})\\ \nonumber &=&  (\vec{\pi} \times \vec{\pi})-\frac{i e \kappa}{m}\vec{E} \times (-i\vec{\nabla} + e\vec{A}) + \frac{i e \kappa}{m}(-i \vec{\nabla} + e \vec{A}) \times \vec{E}.
	\end{eqnarray}
	To go further we act the above operator on $\psi$
	\begin{eqnarray}
		(\vec{\pi} \times \vec{\pi})\psi= (-i \vec{\nabla} + e \vec{A}) \times (-i \nabla +e \vec{A}) \psi
		= -i e \, (\vec{\nabla} \times \vec{A}) \, \psi
		= -i e \, \vec{B} \, \psi
	\end{eqnarray}
	and for the second term
	\begin{eqnarray}
		&& \bigg(\vec{E} \times (-i\vec{\nabla} + e\vec{A}) + (i \vec{\nabla} -e \vec{A}) \times \vec{E}\bigg)\psi=
		- i \vec{E} \times (\vec{\nabla} \psi) + 2 e \, \vec{E} \times \vec{A} \, \psi+i \, (\vec{\nabla} \times (\vec{E} \psi)) \\ \nonumber
		&&=\left(- 2e \, (\vec{A} \times \vec{E}) + i \, (\vec{\nabla} \times \vec{E}) 
		- 2i \, \vec{E} \times (\vec{\nabla}) \right) \psi
		= \left( 2 \, \vec{E} \times \vec{\pi} + i \, (\vec{\nabla} \times \vec{E}) \right) \psi
	\end{eqnarray}
	and at the end, we have
	\begin{eqnarray}
		i \, \vec{\sigma} \cdot (\vec{F}_{+} \times \vec{F}_{-})
		= \vec{\sigma} \cdot \left( e \vec{B} 
		+ \frac{e \kappa}{m} \,( 2 \, \vec{E} \times \vec{\pi}
		+ i \, (\vec{\nabla} \times \vec{E})) \right).
	\end{eqnarray}
	Since we only need to calculate the Hamiltonian to the second order we can keep terms to the zeroth order in $ \bigg[ \mathcal{G},  [\mathcal{G}, \mathcal{E} ] + i \, \dot{\mathcal{G}} \bigg]$ and to the first order in $\beta \mathcal{G}^2$.
	\\
	So the $\mathcal{G}^2$ term to our desired order would be:
	\begin{eqnarray}
		[ \mathcal{G}^2 ]^{1 1} = \pi^2 + \frac{e \kappa}{m} \, (\vec{\nabla} \cdot \vec{E})
		+ e \, \vec{\sigma} \cdot \vec{B}
		+ \frac{e \kappa}{m} \, \vec{\sigma} \cdot 
		\left( 2 \, \vec{E} \times \vec{\pi} + i \, \vec{\nabla} \times \vec{E} \right)
	\end{eqnarray}
	and for calculating the  $ \bigg[ \mathcal{G},  [\mathcal{G}, \mathcal{E} ] + i \, \dot{\mathcal{G}} \bigg]$ we would have:
	
	\begin{eqnarray}
		i \, \dot{\mathcal{G}} = i \, \vec{\alpha} \cdot (e\dot{\vec{A}})
	\end{eqnarray}
	and
	\begin{eqnarray}
		[\mathcal{G}, \mathcal{E} ]=[ \vec{\alpha} \cdot \vec{\pi} , \; -e \varphi ] 
		= \left( \vec{\alpha} \cdot (\vec{\pi} \, e \varphi) \right) 
		- e \varphi \left( \vec{\alpha} \cdot \vec{\pi} \right)
		= \begin{pmatrix}
			0 &  i e \, \vec{\sigma} \cdot (\vec{\nabla} \, \varphi) \\
			i e \, \vec{\sigma} \cdot (\vec{\nabla} \varphi) & 0
		\end{pmatrix}
	\end{eqnarray}
	so 
	\begin{eqnarray}
		[ \mathcal{G}, \mathcal{E} ] + i \, \dot{\mathcal{G}} 
		= i e \, \vec{\alpha} \cdot (\vec{\nabla} \varphi + \dot{\vec{A}})
		= -i e \, \vec{\alpha} \cdot \vec{E}
	\end{eqnarray}
	and for the final term we would need to calculate
	\begin{eqnarray}
		[ \vec{\alpha} \cdot \vec{\pi} , \; -i e \, \vec{\alpha} \cdot \vec{E} ]
		= -i e \, \begin{pmatrix}
			[ \vec{\sigma} \cdot \vec{\pi} , \; \vec{\sigma} \cdot \vec{E} ] & 0 \\
			0 & [ \vec{\sigma} \cdot \vec{\pi} , \; \vec{\sigma} \cdot \vec{E} ]
		\end{pmatrix}
	\end{eqnarray}
	and by using the relation;
	\begin{eqnarray}
		(\sigma_i \, \sigma_j) \, (\pi_i \, E_j) 
		= (\vec{\pi} \cdot \vec{E}) + i \, \vec{\sigma} \cdot (\vec{\pi} \times \vec{E})
	\end{eqnarray}
	we would have
	\begin{eqnarray}
		\left( (\vec{\pi} \cdot \vec{E}) - \vec{E} \cdot \vec{\pi} \right) \psi
		= -i \left( \vec{\nabla} \cdot \vec{E} \right) \psi
	\end{eqnarray}
	and
	\begin{eqnarray}
		\left( \vec{\pi} \times \vec{E} - \vec{E} \times \vec{\pi} \right) \psi
		= \left( 2 e \, (\vec{E} \times \vec{A}) - i \, (\vec{\nabla} \times \vec{E}) 
		+ 2 i \, (\vec{E} \times \vec{\nabla}) \right) \psi
		= \left( - 2 \, \vec{E} \times \vec{\pi} 
		- i \, (\vec{\nabla} \times \vec{E}) \right) \psi
	\end{eqnarray}
	which gives us the final term as
	\begin{eqnarray}
		[ \vec{\alpha} \cdot \vec{\pi} ,\, -i e \, \vec{\alpha} \cdot \vec{E} ]
		= e \, \vec{\sigma} \cdot (-i(\vec{\nabla} \times \vec{E}) 
		- 2 e \, (\vec{E} \times \vec{\pi}))-e (\vec{\nabla} \cdot \vec{E}).
	\end{eqnarray}
	So based on these calculations the Pauli equation to the order of $\frac{1}{m^2}$ would be:
	
	\begin{eqnarray}
		i \, \frac{\partial \chi}{\partial t}
		= \left( m + \left( -e \varphi + \frac{e \kappa}{m} 
		\vec{\sigma} \cdot \vec{B} \right) \right) \chi
	\end{eqnarray}
	\begin{eqnarray*}
		+ \frac{1}{2 m} \left(\pi^2 
		+ \frac{e \kappa}{m} (\vec{\nabla} \cdot \vec{E})
		+ e \, (\vec{\sigma} \cdot \vec{B})
		+ \frac{e \kappa}{m} \, \vec{\sigma} \cdot (i (\vec{\nabla} \times \vec{E}) 
		+ 2 (\vec{E} \times \vec{\pi}))\right) \chi
	\end{eqnarray*}
	\begin{eqnarray*}
		+ \frac{e}{8 m^2} \left( \vec{\nabla} \cdot \vec{E} 
		+ i\vec{\sigma} \cdot (\vec{\nabla} \times \vec{E}) 
		+ 2 \, \vec{\sigma} \cdot (\vec{E} \times \vec{\pi}) \right) \chi
		+ \mathcal{O} \left( \frac{1}{m^3} \right)
	\end{eqnarray*}
	which is the result that we will use.

	\bibliographystyle{unsrt}
	\bibliography{refspinspin-Revised}

\end{document}